\documentclass[12pt,floats,graphicx,epsfig,euscript,amsmath,amssymb]{revtex4}

\usepackage[english,russian]{babel}

\usepackage[dvips]{graphicx}

\usepackage{epsfig}

\usepackage{mathrsfs}

\DeclareMathAlphabet\EuScript{U}{eus}{m}{n} \SetMathAlphabet\EuScript{bold}{U}{eus}{b}{n}

\def\lapprox{\,\raise0.4ex\hbox{$<$}\kern-0.8em\lower0.7ex\hbox{$\sim$}\,}

\def\gapprox{\,\raise0.4ex\hbox{$>$}\kern-0.8em\lower0.7ex\hbox{$\sim$}\,}

\begin{document}

%\twocolumn [\hsize\textwidth\columnwidth\hsize\csname

%  @twocolumnfalse\endcsname

\title

{\centerline{\Large\bf Optical absorption  in incoherent and coherent states }
\centerline{\Large\bf of a quantum Hall system}

\author{S. Dickmann}

\affiliation{$$Institute of Solid State Physics, Russian
Academy of Sciences, Chernogolovka, 142432 Russia}

%$^{*}$Email: dickmann@issp.ac.ru

%\date{\today}

\begin{abstract}

\vskip 0mm

\noindent In connection with recent studies of extremely long-living spin-cyclotron excitations (actually
magneto-excitons) in a quantum Hall electron gas, we discuss contribution to the light-absorption
related to the presence of a magneto-exciton ensemble in this purely electronic system. Since the weakly interacting
excitations have to obey the Bose-Einstein statistics, one can expect appearance of a coherent state in the ensemble. A
comparative analysis of both incoherent and coherent cases is done. Conditions for a phase transition from the
incoherent state to the coherent one are discussed.


\end{abstract}

\maketitle

The cyclotron spin-flip excitation (CSFE) in the $\nu\!=\!2$ quantum Hall system, being the lowest-energy
one$\,$\cite{ka84,di05,ku05}, has an extremely long lifetime. The latter is theoretically estimated to be up to several
milliseconds \cite{di13}. Actually, as is usually the case in relaxation problems, the time experimentally turns out to
be shorter due to the presence of additional relaxation channels which could hardly be predicted before specific
experimental study. In fact the CSFE relaxation found experimentally in the unpolarized quantum Hall system created in
a GaAs/AlGaAs heterostructure reaches 100 $\mu$s \cite{ku15} that seems to be a record value for a delocalized state
excited in the conduction-band electron system. Such a slow relaxation suggests that ensemble of the excitations
obeying the Bose statistics may experience at sufficiently high CSFE concentration a transition to a coherent state --
Bose-Einstein condensate. Note that both the CSFE creation and the CSFE monitoring are performed by optical methods
\cite{ku05,ku15}. In this connection, it is interesting to study the contribution to the light absorption related to
the CSFE ensemble in the 2DEG. In the present work we perform a comparative analysis of the absorption by the CSFE
ensemble in incoherent and coherent phases. (This also strongly correlates with the light emission if the resonant
reflection technique is used \cite{ku15}.)

The CSFE is a solution of the many-electron Schr\"odinger equation with the $\delta S\!=\!1$ change of the total spin
as compared to the ground state where $S\!=\!0$. In other words, generally, this excitation is a triplet with $S\!=\!1$
and $S_z\!=1,0,-1$. All three components have equidistant energies gapped by the Zeeman value $|g\mu_BB|$. The
lowest-energy component corresponds to $S_z\!=\!1$ because the $g$-factor is negative in the GaAs heterostructures. We
will consider only these $S\!=\!S_z\!=\!1$ magnetoexcitons in our study.  A noticeable concentration of such
excitations, $N/{\cal N}_\phi\lesssim 0.1$ (${\cal N}_\phi$ is the total number of states in the Landau level), can be
achieved experimentally \cite{ku15}. At high concentrations the inter-excitonic (CSFE-CSFE) interaction seems to become
fairly strong.  Yet, in the following we study the exciton ensemble only in the `dilute limit', thus ignoring the
CSFE-CSFE coupling.  Due to the very long CSFE relaxation time we study the exciton ensemble as a metastable system
with a given number of excitons $N$.

First the dependence of the CSFE energy on the 2D momentum ${\bf q}$ in the $\nu\!=\!2$ unpolarized quantum Hall system
was calculated by C. Kallin and B. Halperin \cite{ka84}. The authors studied the problem to within the first order in
small parameter giveb by the ratio of characteristic Coulomb energy $E_{\rm C}$ to cyclotron energy
$\hbar\omega_c^{(e)}$ ($E_{\rm C}=\langle {\varepsilon}_q\rangle\!\lesssim\!e^2/\kappa l_B$, $\kappa$ is the dielectric
constant, $l_B\!=\!\sqrt{c\hbar/eB}$ is the magnetic length). Besides, they considered the ultra two-dimensional limit
in the absence of any disorder. Really the CSFE energy counted off the ground-state level is determined by formula
${E}_{\bf q}=\delta+{\cal E}_{q}$, where
\begin{equation}\label{100}
{\cal E}_{q}=\int_0^{\infty}\!\!\!{ds}\,e^{-s^2/2}
{\cal F}_{ee}(s)\left(1-\frac{s^2}{2}\right)
  \left[1-J_0(sq)\right]
\end{equation}
is the $q$-dispersion (here and everywhere below ${\bf q}$ is measured in $1/l_B$ units; $J_0$ is the Bessel function), and $\delta\!\equiv\!\hbar\omega_c^{(e)}\!-\!|g\mu_BB|\!+\!{\varepsilon}_0$ is the $q\!=\!0$ energy including the cyclotron and Zeeman ones, and the negative Coulomb shift ${\varepsilon}_0$ remaining nonzero even if $q\!\to\!0$. (${\varepsilon}_0$ is calculated in the work \cite{di05} and experimentally measured in \cite{ku05}.) Here ${\cal F}_{ee}(q)=\frac{e^2}{\kappa l_B}\int\!\!\int
dz_1dz_2e^{-q|z_1-z_2|/l_B}|\chi_e(z_1)|^2|\chi_e(z_2)|^2$, where  $\chi_e(z)$ describes the electron size-quantized functions in the quantum well. The CSFE $q$-dispersion for the $\nu\!=\!1$ filling is the same as ${\cal E}_q$ in the $\nu\!=\!2$
case if obtained within the ``single-mode approximation''. So, an example of the calculation \eqref{100} for a certain real system is presented,
e.g., in Ref. \cite{pi92}. It shows a very weak $q$-dispersion: $|{\cal E}_q|\lesssim 0.01E_{\rm C}$ down to $q\sim 1$ ($\sim 1/l_B$ in common units).

The CSFE represents a purely electronic kind of magnetoexciton \cite{go68} where the quantum-mechanical average of distance between positions of a promoted electron and an effective `hole' (vacancy in the spin-down sublevel of the zero Landau level) is equal $\Delta{\bf r}=l_B{\bf q}\times\!{\hat z}$ \cite{ka84}. Thus this excitation possesses electric dipole-momentum
\begin{equation}\label{200}
{\bf d}_{\bf q}=el_B{\bf q}\times\!{\hat z}.
\end{equation}

$\,$

\vspace{2mm}

I. Using the `excitonic representation' technique (see, e.g, Ref. \cite{dizhiku}) we study an incoherent state of the CSFE ensemble:
\begin{equation}\label{1}
|{\rm ini},N\rangle\!=\!{\cal Q}_{{\bf q}_N}^\dag\!\!{\cal Q}_{{\bf q}_{N\!-\!1}}^\dag...{\cal Q}_{{\bf q}_1}^\dag|0\rangle,
\end{equation}
where operator ${\cal Q}_{\bf q}^\dag\!=\!{N_\phi}\!\!{}^{-1/2}\!\sum_p e^{-iq_{x}\!(p\!+\!q_{y}\!/2)}b_p^\dag
a_{p+q_{y}}$ (first used in works \cite{dz83}), acting on the ground state $|0\rangle$, creates a magnetoexciton with 2D momentum ${\bf q}$; $|0\rangle$ denotes the
$\nu\!=\!2$ ground state with a fully occupied zero Landau level; $a_p^\dag$ is the operator creating an
electron on the upper spin sublevel of the  {\em zero} Landau level with spin-down, i.e. antiparallel to the magnetic field, and $b_{p}^\dag$ creates an electron on
the {\em first} Landau level with the spin directed along the magnetic field ($p$-numbers are also measured in $1/l_B$ units). Considering the general case under the condition $N\!\ll\!{\cal N}_\phi$ where all ${\bf q}$'s are assumed to be different,
one can find the squared norm: $\langle N,{\rm ini}\,|\,{\rm ini}, N\rangle\!=\!1\!+O(N/{\cal N}_\phi)$.

The perturbation operator
responsible for the light absorption has the form
\begin{equation}\label{01}
{\hat{\cal A}}\!=\!A\sum_p\!V_p^\dag a_p^\dag,
\end{equation}
where $V_p^\dag$
is creation operator of a valence heavy-hole, and A is a
certain constant. Operator \eqref{01} is uniquely determined by
two features of absorption: (i) only `vertical' electronic transitions are relevant in the case, i.e. the photon generates pare consisting of a valence hole and an $a$-sublevel electron -- both in the same intrinsic $p$-states of their Landau levels; (ii) all $p$-states equiprobably participate in the absorption process. Such properties of the light absorption are related to the condition ${\mathscr L}k_{{\rm photon}\parallel}\!\ll\!1$
where length ${\mathscr L}$ is a characteristic of the electron 2D-density spatial fluctuations  and $k_{{\rm photon}\parallel}$ is the photon wave-vector component parallel to the electron system plane. This condition actually is of met \cite{ku15}. The action of the
${\hat{\cal A}}$ operator on state $|{\rm ini}, N\rangle$ results in the $A\sum_i|f,{\bf q}_i\rangle$ combination of $N$
states:
\begin{equation}\label{2}
|f,{\bf q}_i\rangle\!=\!-\hat{\cal X}_{\bf q}\!\prod_{j\!\neq\!i}\!{\cal Q}^\dag_{{\bf q}_j}|0\rangle.
\end{equation}
Here $\hat{\cal X}_{\bf q}\!=\!{\cal N}_\phi\!{}^{-1/2}\!\sum_pe^{-iq_{x}\!(\!p\!+\!q_{y}\!/2)}V_p^\dag
b_{p+q_{y}}^\dag$ is the exciton operator which, by acting on the ground state  generates the valence hole and the $b$-sublevel electron.
If $N\!\ll\!{\cal N}_\phi$, then neglecting any interaction of ${\cal Q}^\dag_{{\bf q}_j}|0\rangle$ excitons with each other and with the $\hat{\cal X}_{\bf q}|0\rangle$ exciton, we find $|f,{\bf q}_i\rangle$ has a norm approximately equal to unity. Matrix element squared for
transition to the final state $|f,{\bf q}_i\rangle$ is
\begin{equation}\label{02}
|M_i|^2\!=\!\langle{\bf q}_i,f|{\hat{\cal A}}|{\rm
ini},N\rangle^2\approx|A|^2
\end{equation}
The following calculation of the absorption rate represents a procedure of summation over all possible final states
\begin{equation}\label{3}
{\cal R}_{\rm I}=(2\pi/\hbar)\!\sum_i\!|M_i|^2\delta(D_{{\bf q}_i})\approx(2\pi|A|^2\!/\hbar)\!\sum_i\delta(\!D_{{\bf q}_i}\!).
\end{equation}
The used approach is actually single-excitonic, hence
\begin{equation}\label{4}
D_{{\bf q}}=\hbar\omega\!+\!{E}_{{\bf q}}\!\!-\!E_{v\!-e,{\bf q}},
\end{equation}
where $\omega$ is the probing laser-beam frequency, ${E}_{\bf q}$ the total energy of the CSFE ${\cal Q}_{\bf
q}^\dag|0\rangle$, and $E_{v\!-e,{\bf q}}$ the energy of creation of the valence-hole--conduction-electron
magnetoexciton state,
\begin{equation}\label{33}
|v-e,{\bf q}\rangle={{\cal N}_\phi}\!{}^{-1/2}\!\sum_p e^{-iq_{ix}\!(p\!+\!q_{iy}\!/2)}V_p^\dag b_{p+q_{iy}}^\dag|0\rangle\,.
\end{equation}

The following study, i.e., in fact, summation of the $\delta$-functions in Eq. \eqref{3}, becomes impossible without a
certain concretization concerning the initial state \eqref{1}, representing actually the distribution of the ${\bf
q}_i$-numbers over their possible values. This distribution is established and determined by two competing effects: by
thermal diffusion related to interactions with phonons, and by drift motion, where the drift velocity of the
magnetoexciton $\partial E_{\bf q}/\partial{\bf q}$ is determined by two parameters, namely, momentum ${\bf q}$ and
smooth random electric field $\vec{\mathscr E}\!=\!-{{\mbox{\boldmath $\nabla$}} \!\varphi}({\bf r})$ [${\bf
r}\!=\!(x,y)$].  We assume that only drift motion accompanied by cooling due to phonon emission results in
establishment of the initial state \eqref{1}.

First, let us study a domain with linear dimensions smaller than the spatial dispersion parameter $\Lambda$ of the
smooth random potential $\varphi$ but still larger than the magnetic length. (For definiteness we will consider
$\Lambda$ to be the correlation length of the $\varphi$ spatial distribution, and the mean value of the potential is
$\overline{\varphi}\!\equiv\!0$.) Within this domain we use a gradient approximation considering field ${\varphi}({\bf
R})$ as well as gradients ${\mbox{\boldmath $\nabla$}}\!{}_R{\varphi}$ and coordinate ${\bf R}$ as parameters inherent
to the domain (for example, ${\bf R}$ indicates the domain center.) So, within the domain the electrostatic term in the
Hamiltonian is equal to ${\hat \varphi }\!=\!\vec{\mathscr E}({\bf R})\!\sum_i{\bf r}_i$ and, when choosing
$\vec{\mathscr E}\parallel{\hat x}$, in terms of secondary quantization it is presented as
\begin{equation}\label{5}
{\hat \varphi}={\mathscr E}l_B\left[\,\sum_{\sigma=\uparrow,\downarrow}\!({\hat K}_\sigma^\dag+{\hat K}_\sigma)-\hat{ P}_y\right]
\end{equation}
(${\mathscr E}$ denoting $|\vec{\mathscr E}|$), where
\begin{equation}\label{6}
{\hat P}_y=\!\!\sum\limits_{n,p,\sigma=\uparrow/\downarrow}\!\!p\,c^{\dag}_{np\,\sigma}
c_{n p\,\sigma}
\end{equation}
is a component of the Gor'kov-Dzyaloshinsky momentum operator $\hat{\bf P}$ \cite{go68} rewritten in the `generalized'
form valid for purely electronic magnetoexcitons in a quantum Hall system (see Ref. \cite{di-c-m}; in particular,
$\hat{\bf P}{\cal Q}^\dag_{\bf q}|0\rangle\!=\!{\bf q}{\cal Q}^\dag_{\bf q}|0\rangle$). Operator
\begin{equation}\label{7}
{\hat K}^\dag_\sigma=\!\!\sum\limits_{n,p}\!
\sqrt{\frac{n\!+\!1}{2}}\,c^{\dag}_{n\!+\!1 p\sigma}c_{np\,\sigma}
\end{equation}
is the Landau-level `raising' operator; $c^{\dag}_{n p\sigma}$ is the creation operator for the $n$-th Landau level,
e.g., $a_p\!\equiv\!c_{0 p\downarrow}$ and $b_p\!\equiv\!c_{1 p\uparrow}$. Now we can obtain the contribution of
electrostatic term \eqref{5} to magnetoexciton energy $E_{\bf q}$. The first order correction is $\langle 0|{\cal
Q}_{\bf q}|{\hat \varphi}|{\cal Q}_{\bf q}^\dag|0\rangle\!\equiv\!-l_B\vec{\mathscr E}\!\times\!\langle 0|{\cal Q}_{\bf
q}|{\hat{\bf  P}}^\dag|{\cal Q}_{\bf q}^\dag|0\rangle=l_B({\bf q}\times\vec{\mathscr E})_z$. In principle, the second
order correction should be determined by the $K$-terms of operator \eqref{5}. By so calculating, attention should be
given to the fact that the energy of the ${\cal Q}_{\bf q}^\dag|0\rangle$ state is counted off the ground state
energy. However, the latter for its part also includes second order electrostatic correction, and  both
corrections turn out to be equal. Thus, the difference vanishes and the total correction is reduced to a null result
\cite{foot1}.

Now we can write out the relevant energy of the magnetoexciton within the domain. We consider its $q$ dispersion
\eqref{100} in the vicinity of the weakly manifested roton minimum (see Ref. \cite{pi92}). So, to within a constant independent of the parameters ${\bf q}$ and
${\bf R}$, the relevant part of the energy is
\begin{equation}\label{8}
{\cal E}'\!({\bf q},{\bf R})=\alpha(q-q_{01})^2+l_B({\bf q}\!\times\!\vec{\mathscr E})_z\,,
\end{equation}
where parameters $\alpha$ and $q_{01}$ are positive and supposedly estimated as $\alpha\!\simeq\!0.1\,$meV and
$q_{01}\!\simeq\!1$ (in the $1/l_B$ units). This energy reaches a local minimum value at ${\bf q}={\bf q}_m({\bf R})\!=\!-\left(l_B\!/2\alpha\!+\!q_{01}\!/{\mathscr E}\right)\vec{\mathscr
E}\!\!\times\!\!{\hat z}$ which is the root of equation $\partial {\cal
E}/\partial{\bf q}\!=\!0$. The local minimum corresponds to zero velocity and, hence, to the zero drift velocity of the
electron and effective `hole' composing the magnetoexciton.
Physically this means vanishing of the total electric field that acts on each quasiparticle since
%${\bf d}_m\!=\!l_B{\hat z}\!\times\!{\bf q}_m$
the electron-`hole' interaction field just compensates the external one. It
is natural to consider the initial metastable state corresponding to this minimum. Due to cool-down
processes, diffusion and drift, which are fast compared to the CSFE lifetime, the magnetoexciton gets ``stuck'' in the smooth random
potential with energy $\delta'\!+{\cal E}_m({\bf R})$, where
\begin{equation}\label{9}
{\cal E}_m({\bf R})={\cal E}'({\bf q}_m,{\bf R})\!=-l_Bq_{01}{\mathscr E}-({\mathscr E}l_B\!)^2/4\alpha.
\end{equation}
($\delta'$ is a constant independent of ${\bf R}$; cf. the study of spin-exciton kinetics in Ref. \cite{zh14}). Thus,
the system represents a frozen but chaotic state held by the smooth random potential.  Magnetoexciton trapping occurs
only in domains where $|{\cal E}_m|\!\gtrsim\!T$, hence, for $T\!\sim\!1\,$K we get $l_B{\mathscr E}\!\gtrsim\!1\,$K
(i.e. ${\mathscr E}\!\gtrsim\!100\,$V/cm). In principle, this is in agreement with the mean ${\mathscr E}$ value
expected for the wide-thickness quantum well employed in the experiment.

The final state $|v-e,{\bf q}_m\rangle$ emerging within the domain as a result of photon absorption is explicitly
defined. It represents a `common' two-particle 2D magnetoexciton studied, for instance, in Ref. \cite{ler80} which is
now, however,  considered against the background of the zero Landau level completely occupied by conduction-band
electrons. This background is a rigid system whose state, if calculated to the first order in the Coulomb interaction
and external electric field, is not changed even in the presence of the magnetoexiton. The occupied electronic Landau
level for its part  does not influence the $q$-dispersion of the energy $E_{v\!-e,{\bf q}}$ studied within the same
approximation. However, if we discuss the $q$-dispersion using, as in Refs. \cite{go68,ler80}, the parabolic valence band
model, allowance should be made for the fact that the corresponding effective formfactor ${\cal F}_{ve}(q)$ in the
dispersion equation similar to Eq. \eqref{100} is still larger than ${\cal F}_{ee}(q)$. (It occurs owing to the greater
``compactness'' of heavy-hole size-quantized function $|\chi_v(z)|^2$ compared to electron function  $|\chi_e(z)|^2$.) Energy
dependence on the external random field can be studied in the way described above { with the only replacement of
the field operator ${\hat \varphi }$ by the operator} ${\hat \varphi }-\vec{\mathscr E}\!({\bf R})\,{\bf r}_h$, where
${\bf r}_h$ is the position of the valence hole within the domain. The first order correction in electric field
${\mathscr E}$ is the same as for the purely electronic magnetoexciton. Meanwhile the second order correction does not
vanish. Indeed, first, the $\sim {\mathscr E}^2$ corrections to energies of purely electronic states
$b_p^\dag|0\rangle$ and $|0\rangle$ determined by the Eq. \eqref{7} operators do not compensate each other now, unlike the situation above taken place with the states ${\cal Q}^\dag_{\bf q}|0\rangle$ and $|0\rangle$. Second,
the $\sim {\mathscr E}^2$ correction to the state $V_p^\dag|0\rangle$ should also be taken into account. As a result,
the total correction is the same as that found by L.P. Gor'kov and I.E. Dzyaloshinsky \cite{go68}. Finally, the
relevant part of the $E_{v\!-e,{\bf q}_m}$ energy is
\begin{equation}\label{10}
{\cal E}_{v\!-e}({\bf q}_m,{\bf R})=\beta (q_m-q_{02})^2+l_B({\bf q}_m\!\times\!\vec{\mathscr E})_z\displaystyle{-\frac{{\mathscr E}^2l_B^2}{2\hbar}\left(\frac{1}{\omega_{c}^{(e)}}+\frac{1}{\omega_{c}^{(h)}}\right)}\,,
\end{equation}
where $\beta\!>\!\alpha$, $q_{02}\!\sim\! 1$; $\omega_{c}^{(e)}$ and $\omega_{c}^{(h)}$ are the cyclotron frequencies
in the conduction and valence bands, respectively. The last term in Eq. \eqref{10} is definitely small and, within the
framework of the absorption mechanism studied here, it has to be taken into account only in the symmetric case where
$\alpha\approx\beta$ and $q_{01}\approx q_{02}$. However, this term also becomes essential in further calculations relating to the coherent ensemble of magnetoexcitons.

Domains participating in summation \eqref{4} must satisfy two conditions: (i) they contain a magnetoexciton and (ii)
correspond to a vanishing argument of the $\delta$-function. Therefore, {to within a constant} $C$ independent of
the coordinate, we come to the equation $\hbar\omega-C={\cal E}_{v\!-e}({\bf q}_m,{\bf R})-{\cal E}_m({\bf R})$
for ${\bf R}$. This equation, by substituting expressions \eqref{9} and \eqref{10}, can be rewritten as
\begin{equation}\label{11}
F({\bf R})=\hbar\omega-{const}\,,
\end{equation}
where
\begin{equation}\label{12}
F({\bf R})=(\beta/\alpha)l_B(q_{01}\!-q_{02}){\mathscr E}+(\beta-\alpha)({\mathscr E}l_B\!)^2\!/4\alpha^2\,,
\end{equation}
and $const$ is a combination of the forbidden gap and the Coulomb, Zeeman and cyclotron shifts relevant to the case
(the $\propto\!{\mathscr E}^2/\omega_c^{(e,h)}$ terms are ignored). The frequency
$\omega$ of the probing laser beam seems to be appropriately tuned in order to ensure a maximum signal, i.e. specifically, in
terms of our study, to provide the maximum number of domains participating in the sum \eqref{3}. The solution of Eq.
\eqref{11} determines  a certain line in the $(X,Y)$ plane. We introduce local orthogonal coordinates $s$ and $u$,
where $s$ is the length along the line, and unit vector ${\hat u}$ is parallel to ${\mbox{\boldmath $\nabla$}}\!{}F$.
Obviously, in the general case of a smooth and random field $F({\bf R})$ (these features are related to smoothness and
randomicity of ${\mathscr E}$) the line represents a closed non-self-intersecting curve (loop) with a length determined
by the inhomogeneity characteristic  $\Lambda$ and frequency $\omega$. First, consider a certain domain $du\!\times\!ds$ {adjacent to the curve}. If the distribution of magnetoexcitons over area $S\!=\!2\pi l_B^2{\cal N}_\phi$ of the spot
created by the pumping laser is equiprobable, then the probability to find a CSFE within the domain is $(N/S)duds$ and
the contribution of the domain to the sum in Eq. \eqref{3} is $(N/S)ds\!\int\delta(|{\mbox{\boldmath
$\nabla$}}\!{}F|u)du=Nds/S|{\mbox{\boldmath $\nabla$}}\!{}F|$. Summating over all such domains we estimate the
contribution of a single loop to the sum \eqref{3} as $N{\mathscr L}_\omega/S\overline{|{\mbox{\boldmath
$\nabla$}}\!{}F|}$, where ${\mathscr L}_\omega$ is the loop length and $\overline{|{\mbox{\boldmath $\nabla$}}\!{}F|}$
is the  mean value along the loop. It is natural to assume that at frequency $\omega$ corresponding to the maximum of
the absorption signal the {largest} contribution to the \eqref{3} signal is provided by `standard' loops (cf. discussion on electron-drift trajectories in a quantum Hall system presented in Ref. \cite{io96}). For
those we estimate ${\mathscr L}_\omega\sim \pi\Lambda'$ where $\Lambda'$ is the linear characteristic of the
inhomogeneity for the ${\mathscr E}\!=\!|{\mbox{\boldmath $\nabla$}}\!{}\varphi|$ field, and, hence,
$\Lambda'\!\sim\!\Lambda$. A more delicate estimate shows that $\Lambda'\!=k\Lambda$ where $k\!<\!1$. Indeed, for
example, if it is assumed that ${\mathscr E}\simeq\sqrt{\overline{({\mbox{\boldmath $\nabla$}}\!{}\varphi)^2}}$ and
$\varphi$ is a Gaussian random field, then $\overline{({\mbox{\boldmath
$\nabla$}}\!{}\varphi)^2}\!=\!2(\Delta/\Lambda)^2$ [$\Delta$ describing the potential amplitude
$\left(\overline{\varphi^2}\right)^{1/2}$]. Therefore, ${\mathscr E}\simeq\sqrt{2}\Delta/\Lambda$, and
$k\!\simeq\!1/\sqrt{2}$ in this case. By analogy we find $\overline{|{\mbox{\boldmath $\nabla$}}\!{}{\mathscr
E}|}\simeq \Delta/{\Lambda'}^2$ and $\overline{|{\mbox{\boldmath $\nabla$}}\!{}{\mathscr E}^2|}\simeq
\Delta^2/{\Lambda'}^3$. These estimates should be substituted into $\overline{|{\mbox{\boldmath $\nabla$}}\!{}F|}\sim
(\beta/\alpha)l_B|q_{01}\!-q_{02}|\overline{|{\mbox{\boldmath $\nabla$}}\!{}{\mathscr
E}|}\!+\!(\beta\!-\!\alpha)\overline{|{\mbox{\boldmath $\nabla$}}\!{}{\mathscr E}^2|}l_B^2\!/4\alpha^2$. The last step
in performing the summation in Eq. \eqref{3} is the calculation of the total number of standard loops corresponding to
frequency $\omega$. Intuitively, this number is $\gamma S/(\pi{\Lambda'}^2/4)$ with the factor $\gamma\simeq 1/4$ for
standard loops. Multiplying it by the contribution of one standard loop found above, we obtain an estimate for the
absorption rate ${\cal R}_{I}\!=\!{\cal K}_{\rm I}N$, where ${\cal K}_{\rm I}$ is the ``oscillator strength'':
\begin{equation}\label{13}
{\cal K}_{\rm I}=2\pi|A|^2\!/\hbar\Phi\Delta\,.
\end{equation}
In this formula
\begin{equation}\label{133}
\Phi(B,\Lambda,\Delta)\sim|q_{01}-q_{02}|\frac{\sqrt{2}\beta l_B}{\alpha\Lambda}+(\beta/\alpha-1)\frac{\Delta l_B^2}{2\alpha\Lambda^2}\,.
\end{equation}
Since $\alpha$, $\beta$ and $l_B$ are inversely proportional to the square root of the magnetic field $B$, the
oscillator strength  \eqref{13} should grow with $B$.

$\,$

\vspace{2mm}

II. The employed single-exciton approximation fails with growing magnetoexciton density $n\!=\!N/S$. We discuss the
dependence of the oscillator strength on density $n$ at temperature which is certainly assumed  to be lower than the
value that enables to consider the incoherent magnetoexciton system as a frozen spatial chaos.  Two scenarios of the
influence of the inter-exciton interaction on the oscillator strength may be assumed. The first represents a gradual
evolution: the larger is the magnetoexciton density, the smoother becomes the effective random potential since the
increasing density apparently results in larger effective correlation length $\Lambda$. Indeed, dipole momenta $d_{{\bf
q}_m}$ \eqref{200} oriented to minimize electrostatic energy should also create in the 2D space a screening electric
field reducing the external one.  The absorption signal should grow with weakening of the random electric field
${\mathscr E}\sim \Delta/\Lambda$ [see Eqs. \eqref{13} and \eqref{133}].

In the other scenario, which we will consider in more detail, the oscillator strength increases abruptly at a certain
value of $n\!=\!n_c$.  This increase can be explained by spontaneous rearrangement
of the magnetoexciton system. We do not study the origin and features of this `phase transition'
definitely related to the inter-magnetoexciton interaction and favorable for occurrence of a coherent state. Moreover,
we will stay within the framework of a model formally ignoring the inter-magnetoexciton interaction.

We demonstrate how the light absorption rate can be estimated within the framework of the model of the coherent state
where a considerable number of magnetoexcitons belongs to the same state, i.e. they have equal wave vectors. First, we
consider a cluster with area ${\cal L}\!\times\!{\cal L}$ (so that ${\cal N}_\phi\!=\!{\cal L}^2\!/2\pi l_B^2$)  where all $N$
excitons in the cluster form a single coherent state. Now, instead of the initial state \eqref{1}, we have
\begin{equation}\label{120}
|N\rangle\!=\!({\cal Q}_{\bf q}^\dag)^N|0\rangle.
\end{equation}
The energy $E_{N{\bf q}}$ of this state does not depend on any spatial fluctuations of the electrostatic field in the
case of a large size of the cluster: ${\cal L}\gg\Lambda$. Indeed, at constant ${\bf q}$ summation of electrostatic
contributions $l_B({\bf q}\!\times\!{\hat z}){\mbox{\boldmath $\nabla$}}\!\varphi({\bf r})$ over the cluster area is
reduced to integration $\propto\int\!d{\bf r}{\mbox{\boldmath $\nabla$}}\!\varphi({\bf r})$ and thereby yields zero
result.   However, if ${\cal L}\!\lesssim\!\Lambda$, the electrostatic energy still contributes to $E_{N,{\bf q}}$. The
norm of state \eqref{120} is calculated in the same way as it was done earlier in the case of ${\bf q}\!\equiv\!0$
\cite{di96}. The result,
\begin{equation}\label{norm2}
R_N^2\!=\!\langle N|N\rangle\!=\!N!{\cal N}_\phi!/{\cal N}_\phi^N\!(\!{\cal N}_\phi\!-\!N\!)!,
\end{equation}
does not depend on ${\bf q}$. Now we find the result of the $\hat{\cal A}$ operation [see Eq. \eqref{01}] on the
initial state: $\hat{\cal A}|N\rangle=-AN\hat{\cal X}_{\bf q}|N\!-\!1\rangle$, where again $\hat{\cal X}_{\bf
q}\!=\!{\cal N}_\phi^{-1/2}\!\sum_pe^{-iq_{x}\!(\!p\!+\!q_{y}\!/2)}V_p^\dag b_{p+q_{y}}^\dag$.  Within our
approximation ignoring any inter-excitonic coupling we obviously can consider $|f_{\bf q}\rangle\!=\!\hat{\cal X}_{\bf
q}|N\!-\!1\rangle$ as the final state that has norm equal to $R_{N\!-\!1}$ and energy $E_{N\!-\!1\,{\bf
q}}\!+\!{E}_{v-e}$ [see Eq. \eqref{10} for ${\cal E}_{v-e}$], and thus calculate the transition matrix element squared
\begin{equation}\label{Mtr}
|M_N|^2=\left(|\langle f|{\cal A}|N\rangle/R_nR_{N\!-\!1}\right)^2\approx |A|^2N.
\end{equation}
(the $N\!\ll\!{\cal N}_\phi$ condition is used). This result is by factor $N$ larger than $|M_i|^2$ found in the above
calculation (see also \cite{foot2}). However, the comparative absorbing capacity must again be estimated by calculating
the oscillator strength, and again we are forced to take into account the external random field.

Indeed, according to Eq. \eqref{10} the energy of the ${\cal X}_{\bf q}|0\rangle$ exciton depends on the field
${\mathscr E}({\bf R})$. We divide the cluster area ${\cal L}\!\times\!{\cal L}$ into small domains parameterized by
coordinate ${\bf R}$, and therefore consider ${\mathscr E}({\bf R})$ and ${\mbox{\boldmath $\nabla$}}\!{\mathscr
E}\!({\bf R})$ within every domain as constant parameters. (Linear dimension of the domain is assumed to be smaller
than $\Lambda$ but larger than $l_B$.) Then the matrix element for transition resulting in creation of the ${\cal
X}_{\bf q}|0\rangle$ exciton within the ${\bf R}$-domain is determined by Eq. \eqref{Mtr} with  $N$ replaced by
$n\,d{\bf R}$. ($n=N/{\cal L}^2$ to describe density of CSFEs considered to be constant in the cluster.) Thus
the absorbtion rate represents a sum over all domains -- actually integration over the 2D space:
\begin{equation}\label{rate}
{\cal R}_{\rm
II}\!=\!\frac{2\pi |A|^2n}{\hbar}\!\int\!d{\bf R}\,\delta(\hbar\omega\!+\!{E}_{N,{\bf q}}\!\!-{E}_{N\!-\!1\,{\bf q}}\!-E_{v-e,{\bf q}}).
\end{equation}
Considering the difference ${E}_{N{\bf q}}\!\!-{E}_{N\!-\!1\,{\bf q}}\!-{\cal E}_{v-e}({\bf q},{\bf R})$ within the
domain [see Eq. \eqref{10} for ${\cal E}_{v-e}$] we conclude that first-order electrostatic terms $l_B({\bf
q}\!\times\!{\hat z}){\mbox{\boldmath $\nabla$}}\!{\mathscr E}\!({\bf R})$ are again compensating each other in the
initial and final states, and therefore do not enter the difference. Now, however, the electrostatic contribution to
the argument of the $\delta$-function in Eq. \eqref{rate} is related to the $\propto{\mathscr E}^2$ terms in Eq.
\eqref{10}.  The situation differs from the previous one in replacement of the field $F({\bf R})$ with the field
$-{\mathscr E}^2\!({\bf R})\!\left(\!1\!/\omega_{c}^{(e)}\!\!+\!1\!/\omega_{c}^{(h)}\right)\!l_B^2/2\hbar$. Considering
the $(u,v)$ local coordinate system ($d{\bf R}\!=\!dudv$) we choose unit vector ${\hat u}$ directed along the gradient
${\mbox{\boldmath $\nabla$}}\!{\mathscr E}^2\!({\bf R})$. As a result, estimating $|{\mbox{\boldmath
$\nabla$}}{\mathscr E}^2|\sim \Delta^2/{\Lambda'}^3$, we find according to Eq. \eqref{rate}: first, the contribution of
one `standard' loop to the absorption rate; and, finally, multiplying by the number of standard loops within the
cluster $\sim\!{\cal L}^2/\pi\Lambda'^2$ (assuming ${\cal L}\!\gg\!\Lambda$), the contribution of the cluster to the
absorption rate ${\cal R}_{\rm II}={\cal K}_{\rm II}N$, where the oscillator strength is
\begin{equation}\label{121}
{\cal K}_{\rm II}\sim |A|^2\pi\omega_{c}^{(h)}(\Lambda/l_B\Delta)^2
\end{equation}
(it is taken into account that $\omega_{c}^{(h)}\!<\!\omega_{c}^{(e)}$ and $\Lambda'\!\simeq\!\Lambda$). Thus, the
enhancement of the absorption/reflection signal due to magnetoexciton clustering is ${\cal K}_{\rm II}/{\cal K}_{\rm
I}$. In the case where the first term in Eq. \eqref{133} is assumed to be larger than the second one we obtain
\begin{equation}\label{122}
{\cal K}_{\rm II}/{\cal K}_{\rm I}\sim\frac{\hbar\omega_{c}^{(h)}\!\beta\Lambda|q_{01}-q_{02}|}{\alpha l_B\Delta}\,.
\end{equation}
Using actual experimental data \cite{ku15}: $\Delta/\hbar\omega_{c}^{(h)}\!\simeq\!0.1$, $\alpha/\beta\!\simeq\!0.5$,
$l_B/\Lambda\!\simeq\!0.2$, and $|q_{01}-q_{02}|\!\simeq\!0.1$, we obtain an estimate $${\cal K}_{\rm II}/{\cal K}_{\rm
I}\simeq 10\,.$$

So, when studying the light absorption/refelection, we expect an amplification effect approximately by an order of
magnitude in the case of a quantum transition from the incoherent phase of the CSFE ensemble to the coherent one.

$\,$

\vspace{1mm}

III. In conclusion, we estimate the CSFE concentration $N/{\cal N}_\phi$ at which the above single-exciton approximation
definitely fails.  The effect of magnetoexciton interaction  is studied using the classical approach and considering
the electro-dipole-dipole interaction. In case all their momenta are equal to ${\bf q}$, the total interaction energy of the magnetoexcitons is
\begin{equation}\label{14}
U=\frac{E_{\rm C}l_B^3}{2}\sum_{i,j}\left[{{\bf d}_{\bf q}}\!{}^2\!/R_{ij}^{3}-3({\bf d}_{\bf q}{\bf R}_{ij})^2\!/R_{ij}^5\right]\,,
\end{equation}
where ${\bf d}_{\bf q}$ is given by Eq. \eqref{200}, ${\bf R}_{ij}\!=\!{\bf R}_{i}\!-\!{\bf R}_{j}$ is vector in the 2D space directed from the $i$-th magnetoexciton to the $j$-th one, and  $E_{\rm
C}\!=\!e^2\!/\kappa l_B\approx 9\,$meV if $B\simeq 5\,$T. After averaging over angles between ${\bf q}$ and ${\bf R}_{ij}$ at given
value ${ q}$ and at fixed distance  ${R}_{ij}$ one finds that the average interaction is attractive. This property of the CSFE-CSFE
interaction reveals that (i) in the absence
of disorder the dilute limit is hardly valid for a long-living magnetoexciton ensemble; (ii) in the presence of a smooth random potential the inter-magnetoexcitonic interaction favors
clasterization of magnetoexcitons and thus formation of a coherent phase.

Then changing in Eq. \eqref{14} from summation to integration: $\sum_{i,j}...\!\to (N^2/S)\!\int\!...\,d{\bf R}$ (where ${\bf
R}_{ij}\!\to\!{\bf R}$ and $S={\cal L}^2$ is the area), we obtain an estimate $U\!\sim\!-\pi q^2 l_B^2N^2E_{\rm C}/2lS\equiv -N^2q^2l_BE_{\rm
C}/4l{\cal N}_\phi$. Here $l$ is the lower limit of the integral which has been set equal to the magnetoexciton characteristic `dimension'
$ql_B$ if $q\!\sim\!1$ (however one must consider $l\!\sim\!l_B$ in the $q\!\ll\!1$ case). The interaction energy per one magnetoexciton, $U/N$, should obviously be compared to energy
\eqref{9} holding the magnetoexciton by the external random potential. If $q\!\simeq\!1$, ${\mathscr
E}\!\simeq\!\Delta/\Lambda$ where $\Delta\!\simeq\!0.5\!-\!0.8\,$meV and $\Lambda\!=\!50\,$nm, then the comparison
leads to an estimate for the critical concentration $N/{\cal N}_\phi\!\simeq\,$5-7\%
essentially rearranging the initial incoherent state.

The research was supported by the Russian Science Foundation: grant \#18-12-00246.
The author thanks I.V. Kukushkin and L.V. Kulik for the useful discussion, and is grateful for hospitality of the International Institute of Physics (Natal, Brazil) where considerable part of the work was done.


\begin{thebibliography}{99}
\bibitem{foot1}
This feature is due to the symmetry of the  purely electronic system considered. Formally, as applied to the case of a common two-particle magnetoexciton \cite{go68},
this symmetry would correspond to electron and hole masses satisfying the $m_e=-m_h$ equivalence and thus results in
vanishing of the $\sim\!{\mathscr E}^2$ term in the magnetoexciton energy.
\end{thebibliography}
\end{document}